\documentclass[]{aa}
\usepackage{psfig}
\usepackage{txfonts}
\begin{document}

\title{Resolution requirements for simulating gravitational fragmentation using SPH}

\author{D.\,A.\,Hubber, S.\,P.\,Goodwin \and A.\,P.\,Whitworth}

\offprints{David.Hubber@astro.cf.ac.uk}

\institute{School of Physics \& Astronomy, Cardiff University,
5 The Parade, Cardiff, CF24 3YB, Wales, UK}

\authorrunning{D.A. Hubber, S.P. Goodwin, A.P. Whitworth}

\titlerunning{Simulating gravitational fragmentation using SPH}

\date{08/12/05}

\abstract{Jeans showed analytically that, in an infinite uniform-density isothermal 
gas, plane-wave perturbations collapse to dense sheets if their wavelength, $\lambda$, 
satisfies $\lambda > \lambda_{_{\rm JEANS}} = \left(\pi a^2 / G \rho_{_0}\right)^{1/2}$ 
(where $a$ is the isothermal sound speed and $\rho_{_0}$ is the unperturbed density); 
in contrast, perturbations with smaller $\lambda$ oscillate about the uniform density 
state. Here we show that Smoothed Particle Hydrodynamics reproduces these results 
well, even when the diameters of the SPH particles are twice the wavelength of 
the perturbation. Our simulations are performed in 3-D with initially settled (i.e. 
non-crystalline) distributions of particles. Therefore there exists the seed noise for 
artificial fragmentation, but it does not occur.  We conclude that, although there may 
be -- as with any numerical scheme -- `skeletons in the SPH cupboard', a propensity to 
fragment artificially is evidently not one of them.
}

\maketitle


\section{Introduction} \label{SEC:INTRO}

Stars form through the collapse and fragmentation of molecular clouds. This is a 
highly chaotic and non-linear process; it involves many different physical effects, 
of which the dominant one is arguably gravitational fragmentation; and it involves 
a very large dynamic range of physical scales and complex geometries. Because the 
process is chaotic and non-linear, numerical simulations have a central role to play 
in understanding the interplay between the different physical effects. Because 
gravitational fragmentation is a critical effect, it is essential that numerical 
schemes are able to capture this effect properly, i.e. that true gravitational 
fragmentation is not suppressed by inadequate resolution, and that artificial 
fragmentation does not occur. Because the process involves a large dynamic range 
of physical scales and complex geometries, Smoothed Particle Hydrodynamics has 
been used extensively to simulate star formation (e.g. in 2004 alone, 
Bonnell, Vine \& 
Bate, 2004; Clark \& Bonnell, 2004; 
Delgado-Donate, Clarke \& Bate, 2004a; 
Delgado-Donate et al., 2004b; 
Goodwin, Whitworth \& Ward-Thompson, 2004a,b,c; 
Hennebelle et al., 2004; 
Hosking \& Whitworth, 2004a,b; 
Jappsen \& Klessen, 2004; 
Li, Mac Low \& Klessen, 2004; 
Rice et al., 2004; 
Kurosawa et al., 2004; 
Price \& Monaghan, 2004a,b; 
Schmeja \& Klessen, 2004; 
Whitehouse \& Bate, 2004).

In this paper we present a new demonstration of the ability of Smoothed 
Particle Hydrodynamics to simulate gravitational fragmentation properly, 
even at very poor resolution, using the plane-wave analysis initially 
performed by Jeans (1929). In Section \ref{SEC:SPH}, we briefly describe 
SPH in general, and the implementation we use in particular. In Section 
\ref{SEC:RESO}, we define -- from first principles -- the resolution required 
for simulating gravitational fragmentation, i.e. the so-called Jeans 
Condition. In Section \ref{SEC:TEST}, we describe the Jeans Test, and in 
Section \ref{SEC:ICS}, we explain how the initial conditions for the test 
are set up. In Section \ref{SEC:RESU}, we present the results of the test, 
emphasising how poor the resolution can be made before the results are 
significantly corrupted. In Section \ref{SEC:SGCO}, we derive and 
demonstrate a correction term for use when the Jeans condition is not 
satisfied or only weakly so. In Section \ref{SEC:CONC}, we summarise our 
main conclusions.


\section{Smoothed Particle Hydrodynamics} \label{SEC:SPH}

Smoothed Particle Hydrodynamics (hereafter SPH) is a Lagrangian, particle-based 
scheme, first formulated by Lucy (1977) and by Gingold \& Monaghan (1977). The 
fluid is represented by an ensemble of particles having masses $m_{_i}$, positions 
${\bf r}_{_i}$, velocities ${\bf v}_{_i}$, internal energies $u_{_i}$ and smoothing 
lengths $h_{_i}$. Apart from its gravity, 
the influence of particle $i$ extends only to radius $r = 2h_{_i}$, 
and is weighted by the smoothing kernel $W(r/h_{_i})$. There need be no grids 
or symmetry constraints, and mass is conserved automatically, so there is no 
need to solve a continuity equation. Local functions of state can be evaluated 
at an arbitrary position ${\bf r}$ by summing contributions from all the 
particles $j$ whose smoothing kernel overlaps ${\bf r}$, weighted by $W(|{\bf r}-
{\bf r}_{_j}|/h_{_j})$. During the evolution, ${\bf v}_{_i}$ is updated using 
a sum of terms representing hydrostatic, viscous, gravitational and magnetic 
accelerations. Similarly $u_{_i}$ can be updated using a sum of terms representing 
adiabatic, viscous and radiative `heating'. In principle, the radiative heating 
can be coupled to a treatment of radiation transport, but in practise this 
is normally very computationally expensive, and often a barotropic equation of 
state is used instead.

The SPH code we use here is {\sc dragon}. This is an extensively tested 
code (Goodwin, Whitworth \& Ward-Thompson, 2003a), which uses an octal spatial 
decomposition tree (Barnes \& Hut, 1986) to speed up the calculation of 
gravitational accelerations and to find lists of neighbours. Gravity is 
kernel-softened. Particle smoothing lengths, $h_{_i}$ are adapted to give 
${\cal N}_{_{\rm NEIB}} = 50$ neighbours. A 2nd-order Runge-Kutta 
time-integration scheme is used, with multiple particle time-steps regulated 
by a Courant-like condition. The code uses time-dependent viscosity with 
$\alpha^\star_{_{\rm VISC}} = 0.1$ (Morris \& Monaghan 1997); the option 
exists to reduce the effective shear viscosity still further using the Balsara 
switch (Balsara 1995), but this option is not exercised here. Periodic boundary 
conditions can be imposed. If, as here, self-gravity is involved, the Ewald 
method is used (Hernquist et al., 1991; Klessen, 1997). Since we are imposing 
an isothermal equation of state (i.e. $P = a^2 \rho$) the energy equation 
need not be solved.


\section{Resolution requirements for gravitational fragmentation} \label{SEC:RESO}

The resolution requirements for numerical simulations of gravitational 
fragmentation were first addressed systematically by Truelove et al. (1997) 
and Bate \& Burkert (1977). Truelove et al. (1997) used an Adaptive Mesh 
Refinement (AMR) Finite Difference (FD) code to show that grid-based 
codes must use cell sizes $d$ satisfying a Jeans Condition of the form 
$d < \lambda_{_{\rm JEANS}}/4$, where $\lambda_{_{\rm JEANS}} = 
(\pi a^2 / G \rho)^{1/2}$ is the local Jeans length, $a$ is the local 
isothermal sound speed, and $\rho$ is the local density. If this Jeans 
Condition is not met in FD simulations, artificial fragmentation can occur, 
particularly if artificial viscosity is used.

In this context, the configuration initially explored by Boss \& Bodenheimer 
(1979) has acquired the status of a standard test, since it is believed that 
the simulations of this configuration presented by Truelove et al. (1998) 
achieved convergence. The initial configuration involves a spherical cloud 
with mass $M = 1 {\rm M}_{_\odot}$, radius $R = 5 \times 10^{16}\,{\rm cm}$, 
uniform density $\rho = 3.8 \times 10^{-18}\,{\rm g}\,{\rm cm}^{-3}$, 
uniform temperature $T = 10\,{\rm K}$ and uniform angular speed $\Omega = 
7.2 \times 10^{-13}$ (hence ratio of thermal to gravitational energy 
$\alpha = 0.26$ and ratio of rotational to gravitational energy $\beta 
= 0.16$). An azimuthal $m=2$ density perturbation with fractional 
amplitude $A=0.1$ is then imposed and the subsequent isothermal evolution 
is followed. In the Truelove et al. (1998) AMR simulation of this 
configuration, a binary system forms, with a filament between the two 
components, and -- as predicted by Inutsuka \& Miyama (1992) -- the filament 
does not fragment, but rather collapses to a singularity. This is in direct 
contrast with the earlier FD simulations of the Boss \& Bodenheimer 
configuration reported by Burkert \& Bodenheimer (1993). They used nested 
grids to increase the central resolution, but did not satisfy the Jeans 
Condition, and this resulted in fragmentation of the filament (into 
regularly spaced condensations with planetary masses).

The corresponding Jeans Condition for SPH has been derived by Bate \& Burkert 
(1997), who show that SPH models gravitational fragmentation properly 
(i.e. artificial fragmentation is avoided and true fragmentation captured) 
provided (i) the gravity softening and particle smoothing have similar 
scales (this is achieved automatically here with kernel-softened gravity), 
and (ii) the local Jeans mass is resolved at all times. They suggest that 
the minimum mass that can be resolved by SPH is given by $M_{_{\rm MIN}} 
= 2 {\cal N}_{_{\rm NEIB}} m$, where ${\cal N}_{_{\rm NEIB}} \sim 50$ is 
the mean number of neighbours and $m$ is the mass of a single SPH particle 
(here assumed to be universal). In a subsequent paper (Bate, Bonnell \& 
Bromm, 2002), this has been revised down to $M_{_{\rm MIN}} = 1.5 
{\cal N}_{_{\rm NEIB}} m$, and therefore we have elected to put $M_{_{\rm MIN}} 
= {\cal N}_{_{\rm NEIB}} m$ and to use the factor by which $M_{_{\rm JEANS}}$ 
exceeds $M_{_{\rm MIN}}$ as a measure of the accuracy of the code, 
Requiring 
\begin{eqnarray} \nonumber
M_{_{\rm MIN}} \;=\; {\cal N}_{_{\rm NEIB}}\,m \;\leq\; M_{_{\rm JEANS}} & = & 
\frac{4\,\pi\,\left(\lambda_{_{\rm JEANS}}/2\right)^3\,\rho}{3} \;=\; 
\frac{\pi^{5/2}\,a^3}{6\,G^{3/2}\,\rho^{1/2}} \\
\end{eqnarray}
then reduces to
\begin{eqnarray}
\rho & \leq & \left(\frac{\pi\,a^2}{G}\right)^3\,\left(\frac{\pi}
{6\,{\cal N}_{_{\rm NEIB}}\,m}\right)^2 \,,
\end{eqnarray}
or equivalently
\begin{eqnarray} \label{EQN:JEANSCOND1}
\lambda_{_{\rm JEANS}} & \geq & 4\,h \,,
\end{eqnarray}
i.e. the Jeans length should exceed the diameter of an SPH particle.

Whitworth (1998) has shown analytically that with a standard smoothing kernel, 
and a gravitational softening length comparable to the kernel smoothing 
length, the only gravitational condensations which can form in SPH must (a) 
be genuinely gravitational unstable, and (b) be -- at least approximately -- 
resolved. Thus, failing to satisfy the Jeans Condition simply suppresses true 
fragmentation, rather than promoting artificial fragmentation. The present paper 
confirms these results.

Kitsionas \& Whitworth (2002) have shown that standard SPH simulates the 
Boss \& Bodenheimer test as well as AMR, provided sufficient particles are 
used to satisfy the Jeans Condition.  Moreover, the number of particles can 
be greatly reduced (and with it the amount of memory and computation required) 
by implementing on-the-fly particle splitting. Despite the transient 
high spatial-frequency noise introduced by particle splitting, the filament 
between the two binary components shows no sign of fragmenting, even when 
the simulation is evolved to densities almost 100 times higher than Truelove 
et al. (1998) achieved, and the computational cost is greatly reduced. Thus, 
$\sim 600,000$ particles are required for standard SPH to follow this 
test with an isothermal equation of state to $\rho = 6 \times 10^{-8}\,
{\rm g}\,{\rm cm}^{-3}$ (a more stringent test than attempted by Truelove 
et al., 1998). With particle splitting this can be achieved with $\sim 
45,000$ particles initially, and fewer than $\sim 145,000$ at the end, 
with huge savings in the required CPU.


\section{The Jeans Test} \label{SEC:TEST}

Although the Boss \& Bodenheimer test is a demanding one, it does not have 
an analytic solution, established beyond all reasonable doubt, and therefore 
it is appropriate to explore a test which does have an analytic solution. 

Consider a static infinite medium, with uniform density $\rho_{_0}$, and 
uniform and constant isothermal sound speed $a$, and assume that, even 
though it is self-gravitating, it is in equilibrium. This assumption is 
usually referred to as the `Jeans swindle', since the medium cannot strictly 
be in self-gravitating equilibrium if the density is finite. However, in the 
simulations presented below the uniform density unperturbed state is 
effectively in equilibrium, in the sense that it can be evolved for a long 
time without changing.

Now suppose that the medium is perturbed so that $\rho_{_0} \rightarrow 
\rho_{_0} + \rho_{_1}$ and ${\bf v}_{_0} = {\bf 0} \rightarrow 
{\bf v}_{_0} + {\bf v}_{_1} = {\bf v}_{_1}$. The continuity, Euler and 
Poisson equations reduce to the forms 
\begin{eqnarray} \label{EQN:CONT1}
\frac{\partial \rho_{_1}}{\partial t} & = & -\,\rho_{_0} \nabla \cdot {\bf v}_{_1} 
\\ \label{EQN:EULER1}
\frac{\partial {\bf v}_{_1}}{\partial t} &= &-\,\frac{a^2 \, \nabla 
\rho_{_1}}{\rho_{_0}} \,-\, \nabla \phi_{_1}  \\ \label{EQN:POISSON1}
\nabla^2 \phi_{_1} &= &4\,\pi\,G\,\rho_{_1} 
\end{eqnarray}where $\phi_{_1}$ is the gravitational potential due to the 
perturbed density (e.g. Binney \& Tremaine, 1987). Eliminating ${\bf v}_{_1}$ 
and $\phi_{_1}$ from Eqns. (\ref{EQN:CONT1}) to (\ref{EQN:POISSON1}) then yields
\begin{eqnarray} \label{EQN:WAVE1}
\frac{\partial^2\rho_{_1}}{\partial t^2} \,-\, a^2\,\nabla^2 \rho_{_1}
\,-\, 4\,\pi\,G\,\rho_{_0}\,\rho_{_1} \;=\; 0 \,.
\end{eqnarray}
Substituting a plane wave of the form $\rho_{_1}({\bf r},t) = 
A\,\rho_{_0}\,e^{i(kx\pm\omega t)}$ in Eqn. (\ref{EQN:WAVE1}) gives the dispersion 
relation
\begin{eqnarray} \label{EQN:DISPERSION}
\omega_{_k}^2 \;=\; a^2k^2 \,-\, 4\,\pi\,G\,\rho_{_0} \,.
\end{eqnarray}
Hence we can identify a critical Jeans wave-number, 
\begin{eqnarray} \label{EQN:kJEANS}
k_{_{\rm JEANS}} & = & \frac{(4 \pi G \rho_{_0})^{1/2}}{a} \,,
\end{eqnarray}
and correspondingly a critical Jeans wavelength, 
\begin{eqnarray} \label{EQN:lambdaJEANS}
\lambda_{_{\rm JEANS}} & = & \frac{2\,\pi}{k_{_{\rm JEANS}}} \;=\; 
\left(\frac{\pi\,a^2}{G\,\rho_{_0}}\right)^{1/2} \,,
\end{eqnarray}
separating short wavelength perturbations (which oscillate) 
from long wavelength perturbations (which grow).

To set up an initially stationary plane-wave perturbation, we superimpose two plane 
waves of equal amplitude and wavelength, travelling in opposite directions:
\begin{eqnarray}\label{EQN:DENSPERT}
\rho_{_1}({\bf r},t) & = & \frac{A\,\rho_{_0}}{2}\,\left\{ e^{i(kx-\omega_{_k}t)} \,+\, 
e^{i(kx+\omega_{_k}t)} \right\} \,, \\ \label{EQN:VELOPERT}
{\bf v}_{_1}({\bf r},t) & = & \frac{A\,\omega}{2\,k}\,\left\{ e^{i(kx-\omega_{_k}t)} \,-\, 
e^{i(kx+\omega_{_k}t)} \right\} \, \hat{\bf e}_{_x} \,.
\end{eqnarray}

For short wavelength perturbations ($\lambda < \lambda_{_{\rm JEANS}}$, 
$k > k_{_{\rm JEANS}}$), the dispersion relation (Eqn. \ref{EQN:DISPERSION}) 
indicates that $\omega^2$ is positive, and therefore (switching from $k$ 
to $\lambda$),
\begin{eqnarray} \label{EQN:OMEGOSCI}
\omega_{_\lambda} & = & 2\,\pi\,a\,\left( \frac{1}{\lambda^2} 
\,-\, \frac{1}{\lambda_{_{\rm JEANS}}^2} \right)^{1/2} \,
\end{eqnarray}
is real and the perturbation oscillates. Taking the real parts of Eqns. 
(\ref{EQN:DENSPERT}) and (\ref{EQN:VELOPERT}),
\begin{eqnarray} \label{EQN:DENSOSCI}
\rho_{_1}({\bf r},t) & = & A\,\rho_{_0}\,\cos\left(\frac{2\,\pi\,x}{\lambda}\right)\,
\cos\left(\omega_{_\lambda}t\right) \,, \\ \label{EQN:VELOOSCI}
{\bf v}_{_1}({\bf r},t) & = & \frac{A\,\omega_{_\lambda}}{k}\,\sin\left(\frac{2\,\pi\,x}
{\lambda}\right)\,\sin\left(\omega_{_\lambda}t\right)\,\hat{\bf e}_{_x} \,.
\end{eqnarray}
and the oscillation period is
\begin{eqnarray} \label{EQN:PRODOSCI}
T_{_\lambda} & = & \frac{2\,\pi}{\omega_{_\lambda}} \;=\; \left(\frac{\pi}
{G\,\rho_{_0}}\right)^{1/2}\,\frac{\lambda}{(\lambda_{_{\rm JEANS}}^2 - \lambda^2)^{1/2}} \,.
\end{eqnarray}

For long wavelength perturbations ($\lambda > \lambda_{_{\rm JEANS}}$, 
$k < k_{_{\rm JEANS}}$), the dispersion relation (Eqn. \ref{EQN:DISPERSION}) 
indicates that $\omega^2$ is negative, and therefore $\omega_{_\lambda}$ is 
imaginary. Defining
\begin{eqnarray}
\gamma_{_\lambda} & = & 2\,\pi\,a\,\left( \frac{1}{\lambda_{_{\rm JEANS}}^2} 
\,-\, \frac{1}{\lambda^2} \right)^{1/2} \,,
\end{eqnarray}
we can put $\omega_{_\lambda} = i \gamma_{_\lambda}$. Then, taking the real parts 
of Eqns. (\ref{EQN:DENSPERT}) and (\ref{EQN:VELOPERT}), we have
\begin{eqnarray} \label{EQN:DENSGROW}
\rho_{_1}({\bf r},t) & = & A\,\rho_{_0}\,\cos\left(\frac{2\,\pi\,x}{\lambda}\right)\,
\cosh\left(\gamma_{_\lambda}t\right) \,, \\ \label{EQN:VELOGROW}
{\bf v}_{_1}({\bf r},t) & = & \frac{A\,\gamma_{_\lambda}}{k}\,\sin\left(\frac{2\,\pi\,x}
{\lambda}\right)\,\sinh\left(\gamma_{_\lambda}t\right)\,\hat{\bf e}_{_x} \,.
\end{eqnarray}
The time for the perturbed density on the plane $x = 0$ to grow from $A\rho_{_0}$ to 
${\rm cosh}(1)\,A \rho_{_0} = 1.54\,A \rho_{_0}$ is
\begin{eqnarray} \label{EQN:TIMEGROW}
T'_{_\lambda} & = & \frac{1}{\gamma_{_\lambda}} \;=\; \left(\frac{1}{4\,\pi\,G\,\rho_{_0}}
\right)^{1/2}\,\frac{\lambda}{(\lambda^2 - \lambda_{_{\rm JEANS}}^2)^{1/2}} \,.
\end{eqnarray}


\begin{figure} 
\centerline{\psfig{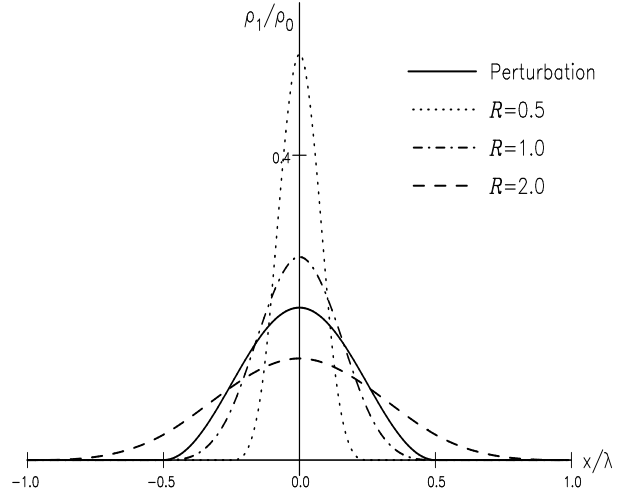}}
\caption{The solid line represents one wavelength of the imposed sinusoidal 
plane-wave perturbation. The other lines represent the smoothing kernels used 
in the results displayed in Fig. \ref{FIG:TIMES}: ${\cal R} = 0.5$ (dotted 
line) the very well resolved case; ${\cal R} = 1.0$ (dash-dot line) the 
marginally resolved case; and ${\cal R} = 2.0$ (dashed line) the under-resolved 
case. The kernels are all scaled so that the integrated area under the kernel 
is equal to the area under the perturbation.}
\label{FIG:KERNELS}
\end{figure}


\begin{figure}
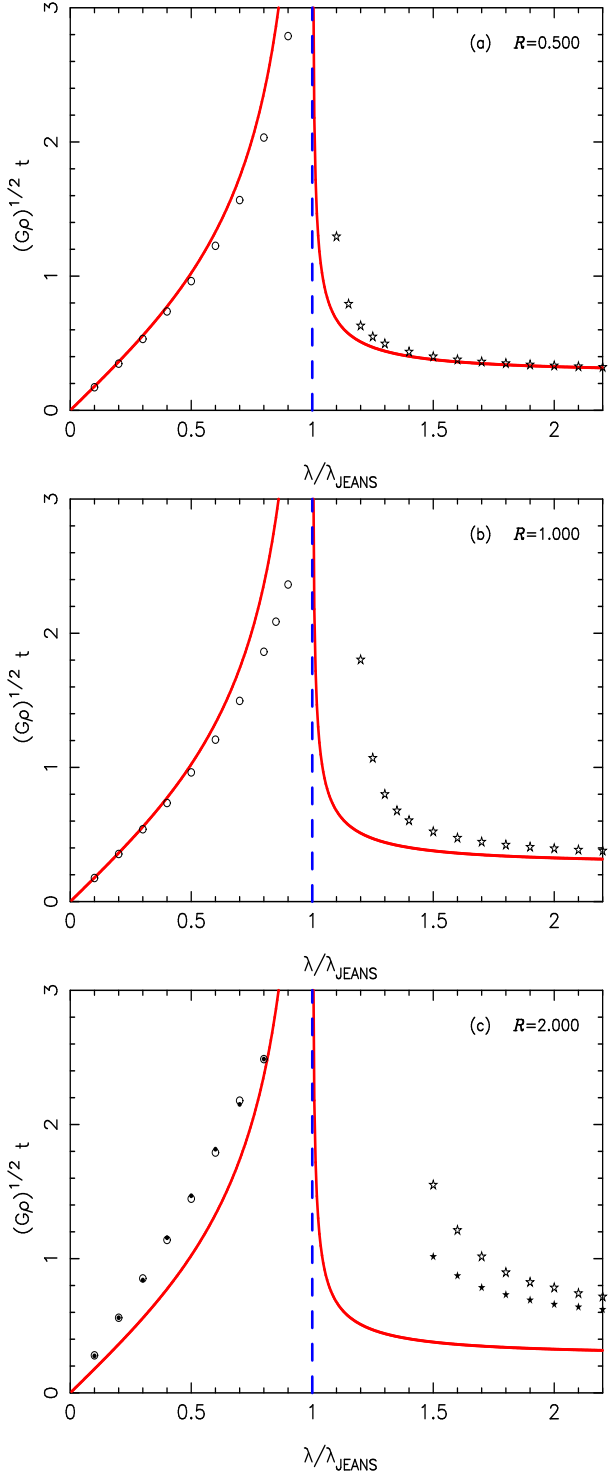

\centerline{\psfig{figure=4100a1.ps,height=6.5cm,width=8cm,angle=270}}
\centerline{\psfig{figure=4100d2.ps,height=6.5cm,width=8cm,angle=270}}
\centerline{\psfig{figure=4100g4-2.ps,height=6.5cm,width=8cm,angle=270}}
\caption{Characteristic timescales for the evolution of plane-wave perturbations, 
as a function of wavelength. The ordinate is the wavelength in units of the 
Jeans length, and the abscissa is the timescale in units of $(G \rho_{_0})^{-1/2}$. 
For perturbations which oscillate (i.e. those with $\lambda < \lambda_{_{\rm JEANS}}$) 
the oscillation period estimated from the SPH simulations is represented by an open 
circle. For perturbations which collapse (i.e. those with $\lambda > 
\lambda_{_{\rm JEANS}}$) the time for the peak density in the SPH simulations to 
increase by a factor 1.54 (see text) is represented by an open star. 
For reference, the analytic timescales are given by solid curves. (a) The very 
well resolved case, ${\cal R} = 0.5$. (b) The marginally resolved case, ${\cal R} 
= 1.0$. (c) The under-resolved case, ${\cal R} = 2.0$. Note that in all cases, 
even the under-resolved one, wavelengths which should oscillate do oscillate 
(i.e. no Jeans-stable perturbations artificially collapse). In (c), the filled circles and 
filled stars represent simulations performed with the correction factor derived in 
Section \ref{SEC:SGCO}.}
\label{FIG:TIMES}
\end{figure}


\section{Initial conditions} \label{SEC:ICS}

Since in both situations (low wavelength perturbations that oscillate and long wavelength 
perturbations that grow) the initial state is
\begin{eqnarray}
\rho({\bf r},t) & = & \rho_{_0}\,\left\{1\,+\,A\,\cos\left(\frac{2\,\pi\,x}{\lambda}
\right)\right\} \,, \\
{\bf v}({\bf r},t) & = & {\bf 0} \,,
\end{eqnarray}
we set up the initial conditions as follows.

First, ${\cal N}_{_{\rm TOTAL}}$ particles are distributed 
randomly within a unit cube and settled using non--self-gravitating SPH and periodic 
boundary conditions. This reduces the Poissonian density fluctuations, and produces an 
approximately uniform, but non-crystalline, density distribution 
(sometimes described as glass-like). 
The mean smoothing length of a particle is given by
\begin{eqnarray}
\bar{h} & = & \left(\frac{3\,{\cal N}_{_{\rm NEIB}}}{32\,\pi\,{\cal N}_{_{\rm TOTAL}}}
\right)^{1/3} \;\simeq\; 1.1427\,{\cal N}_{_{\rm TOTAL}}^{-1/3} \,,
\end{eqnarray}
where we have substituted ${\cal N}_{_{\rm NEIB}} = 50$, and 
the length of the edges of the cube is equal to unity.

Second, a sinusoidal density perturbation is imposed  by adjusting the unperturbed 
$x$-coordinate, $x_{_i}$, of each particle $i$ to a perturbed value, $x'_{_i}$, 
satisfying
\begin{eqnarray} \label{EQN:PERTURBATION}
x'_{_i} \,+\, \frac{A\,\lambda}{2\,\pi} \, \sin\left(\frac{2\,\pi\,x'_{_i}}{\lambda}
\right) & = & x_{_i} \,.
\end{eqnarray}
This equation must be solved numerically for $x'_{_i} = x'(x_{_i})$. We use a 
perturbation with fractional amplitude $A = 0.1$.

Since periodic boundary conditions are being invoked, we can only apply perturbations 
which fit an integer number of wavelengths into the side of the unit cube, i.e.
\begin{eqnarray}
\lambda & = & n_{_\lambda}^{-1} \,,
\end{eqnarray}
where $n_{_\lambda} = 1,\,2,\,3,\,{\rm etc.}$.

A convenient measure of the resolution is the ratio of the 
mean diameter of an SPH particle, $\bar{d} = 4\bar{h}$, to the wavelength of the 
perturbation, $\lambda = n_{_\lambda}^{-1}$, i.e. 
\begin{eqnarray}
{\cal R} \;=\; \frac{\bar{d}}{\lambda} \;=\; n_{_\lambda}\,4\,\bar{h} \;=\; n_{_\lambda}
\,\left(\frac{6\,{\cal N}_{_{\rm NEIB}}}{\pi\,{\cal N}_{_{\rm TOTAL}}}\right) \,.
\end{eqnarray}
Thus a small value of ${\cal R}$ corresponds to good resolution, and the Jeans 
condition (Eqn. \ref{EQN:JEANSCOND1}) can then be rewritten in the form
\begin{eqnarray} \label{EQN:JEANSCOND2}
{\cal R} & \leq & 1 \,.
\end{eqnarray}
where $\lambda = \lambda_{_{\rm JEANS}}$.
The number of SPH particles in one Jeans mass is then
\begin{eqnarray}
\frac{M_{_{\rm JEANS}}}{m} & = & \frac{{\cal N}_{_{\rm NEIB}}}{{\cal R}^3} \,.
\end{eqnarray}

The Jeans wavelength can be varied arbitrarily by changing the isothermal sound speed, 
$a$.


\section{Test results} \label{SEC:RESU}

In Figure \ref{FIG:KERNELS}, we compare the sinusoidal perturbation with the 
smoothing kernel for three values of the resolution: (a) ${\cal R} = 0.5$, i.e. 
very well resolved with $8 {\cal N}_{_{\rm NEIB}} = 400$ SPH particles in one 
Jeans mass; (b) ${\cal R} = 1.0$, marginally resolved, with ${\cal N}_{_{\rm NEIB}} 
= 50$ SPH particles in one Jeans mass; and (c) ${\cal R} = 2.0$, under-resolved, 
with just ${\cal N}_{_{\rm NEIB}}/8 \simeq 6$ SPH particles in one Jeans mass. 
We see why the resolution ${\cal R} = 1.0$ is critical, in the sense that the 
smoothing kernel is closely matched to the perturbation (see Fig. \ref{FIG:KERNELS}), 
and there are just enough SPH particles to resolve a three-dimensional object.

In Figure \ref{FIG:TIMES} we plot the results of SPH simulations of plane-wave 
perturbations having different values of $\lambda / \lambda_{_{\rm JEANS}}$. 
Each panel of Fig. \ref{FIG:TIMES} corresponds to a different resolution, 
${\cal R}$,  viz. (a) ${\cal R} = 0.5$, (b) ${\cal R} = 1.0$, and (c) ${\cal R} 
= 2.0$. If the perturbation oscillates, we plot with an open circle the 
oscillation period. If the perturbation grows, we plot as an open star the time 
required for the amplitude to increase by a factor ${\rm cosh}(1) = 1.54$. The 
analytic predictions for these times (Eqns. \ref{EQN:PRODOSCI} and 
\ref{EQN:TIMEGROW}) are shown as solid lines. From these plots, we can draw 
the following two important conclusions.

(i) There is no artificial fragmentation. Even when the resolution is poor 
(i.e. ${\cal R} = 2.0$, Fig.\ref{FIG:TIMES}(c)), perturbations which should 
oscillate ($\lambda < \lambda_{_{\rm JEANS}}$) do oscillate, and perturbations 
which should grow ($\lambda > \lambda_{_{\rm JEANS}}$) do grow. Furthermore we 
should be mindful that ${\cal R} = 2.0$ corresponds to a strong violation 
of the Jeans Condition, with only $\sim 6$ particles in a Jeans mass. It would 
never be tolerated in a simulation of collapse and fragmentation. 

(ii) With ${\cal R} \stackrel{<}{\sim} 1$ (good to marginal resolution), the 
characteristic timescales are 
reproduced well, except for the cases where $\lambda \sim \lambda_{_{\rm JEANS}}$. 
Marginally Jeans stable perturbations oscillate a little too rapidly, and marginally 
Jeans unstable perturbations grow rather more slowly than they should.

(iii) The effect of large ${\cal R}$ (poor resolution) is to shift the asymptote 
separating stable wavelengths from unstable ones, from $\lambda = \lambda_{_{\rm 
JEANS}}$ to a slightly longer wavelength, i.e. to stabilise marginally Jeans 
unstable wavelengths, as if the temperature had been increased slightly. This is 
the reason for the gap on Fig. \ref{FIG:TIMES}(c) between $\lambda/\lambda_{_{\rm JEANS}} 
= 1.0$ and $\lambda/\lambda_{_{\rm JEANS}} = 1.4$. A perturbation with (say) 
$\lambda = 1.2 \lambda_{_{\rm JEANS}}$ is stabilised and initially oscillates, 
but at the same time, because of the poor resolution, its energy is transferred 
to longer wavelengths modes which then become unstable with a shorter timescale 
than the initial perturbation, and therefore it is not possible to 
evaluate an oscillation period. Thus poor resolution simply suppresses the 
collapse of marginally unstable modes.


\section{Correcting for the self-gravity of an individual SPH particle} \label{SEC:SGCO}

In standard self-gravitating SPH, the mutual gravitational force between two 
different SPH particles is included in the equation of motion, but the self-gravity 
of an individual particle is not. When the Jeans mass is not well resolved, we can 
improve the performance of the code by correcting for the fact that part of the 
pressure of an SPH particle must be used to support the particle against its own 
self gravity , rather than pushing on other particles. To formulate this correction, 
consider particle $i$ in isolation.  If its mass is $m_{_i}$ and its sound speed 
is $a_{_i}$, then
\begin{eqnarray}
3 \int P\,dV & = & 3\,m_{_i}\,a_{_i}^2 \,,
\end{eqnarray}
where the integral is over the volume of the SPH particle. 
We use $a_{_i}$ because in general each SPH particle has its own  
sound speed which differs from that of other particles.
The Virial Theorem 
tells us that, if the particle is to be in hydrostatic equilibrium, this integral 
must equal the magnitude of its self-gravitational potential energy, which is
\begin{eqnarray}
|\Omega| & = & \frac{G\,m_{_i}^2\hat{W}}{h_{_i}} \,.
\end{eqnarray}
Here $h_{_i}$ is its smoothing length, and
\begin{eqnarray}
\hat{W} & = & \int_{s=0}^{s=2}\,\cdot\,\int_{s'=0}^{s'=s}
\,W(s')\,4\,\pi\,s'^2\,ds'\,\cdot\,W(s)\,4\,\pi\,s\,ds 
\end{eqnarray}
is an integral which can be worked out given the dimensionless kernel function $W(s)$. 
It follows that the effective sound speed squared, $a_{_{i,{\rm EFF}}}^2$, should 
be reduced below the true sound speed squared, $a_{_i}^2$, viz  
\begin{eqnarray}
a_{_{i,{\rm EFF}}}^2 = a_{_i}^2 - \frac{G\,m_{_i}\hat{W}}{3\,h_{_i}}\,.
\end{eqnarray}
For the standard M4 kernel which we use, 
\begin{eqnarray} \label{EQN:M4}
W(s) = \frac{1}{\pi}
\left\{ \;\; \begin{array}{ll}
1 - \frac{3}{2}s^2 + \frac{3}{4}s^3 \;\;\;\; & {\rm if}\;\;0 \leq s \leq 1\,; \\
\frac{1}{4}(2-s)^3 \;\;\;\; & {\rm if}\;\;1 \leq s \leq 2\,; \\
0 \;\;\;\; & {\rm if}\;\;s > 2 \,.
\end{array} \right .
\end{eqnarray}
(Monaghan \& Lattanzio, 1985), $\hat{W} = 0.505\,$.

We have repeated the simulations with ${\cal R} = 2.0$, invoking this correction 
factor, and the results are presented as filled circles and filled stars on Fig. 
\ref{FIG:TIMES}(c). As expected, the collapsing wavelengths collapse more rapidly 
(although still more slowly than the analytic solution).


\section{Conclusions} \label{SEC:CONC}

The conclusions are very simple. SPH using the standard M4 kernel and kernel-softened 
gravity (i.e. the standard options) only captures fragmentation which is (a) genuine, 
and (b) resolved. It does not suffer from artificial fragmentation.


\begin{acknowledgements}
DAH gratefully acknowledges the support 
of a PPARC postgraduate studentship, SPG acknowledges the support 
of a UKAFF Fellowship and APW acknowledges the 
support of a European Commission Research Training Network awarded 
under the Fifth Framework (Ref. HPRN-CT-2000-00155). We thank the 
referee, Ralf Klessen, for helpful comments and suggestions.
\end{acknowledgements}



\end{document}